\documentclass[aps,prb,reprint,amsmath,amssymb,superscriptaddress,showpacs,floatfix]{revtex4-1}
\usepackage{graphicx}
\usepackage{dcolumn}
\usepackage{bm}
\usepackage[colorlinks, allcolors=blue]{hyperref}
\usepackage[usenames, dvipsnames]{color}
\usepackage{setspace}
\usepackage{scalerel}
\usepackage{stackengine}
\stackMath
\newsavebox\tmpbox
\newcommand\widefrown[1]{\ThisStyle{%
\sbox\tmpbox{$\SavedStyle#1$}%
\stackon[0pt]{\usebox{\tmpbox}}{%
\stretchto{%
  \scaleto{%
    \scalerel*[\wd\tmpbox]{\mkern-.8mu\frown\mkern-.8mu}%
    {\rule[-\textheight/2]{1ex}{\textheight}}%
  }{\textheight}%
}{0.8ex}}%
}}
\begin{document}

\title{Magnetization plateaus of spin-$\mathbf{\frac{1}{2}}$ system on a 5/7 skewed ladder}

\author{Dayasindhu Dey}
\thanks{First and second authors have contributed equally ~~~~}
\affiliation{Solid State and Structural Chemistry Unit, Indian Institute of Science, Bangalore 560012, India}

\author{Sambunath Das}
\thanks{First and second authors have contributed equally ~~~~}
\affiliation{Solid State and Structural Chemistry Unit, Indian Institute of Science, Bangalore 560012, India}

\author{Manoranjan Kumar}
\email{manoranjan.kumar@bose.res.in}
\affiliation{S. N. Bose National Centre for Basic Sciences, Block - JD, Sector - III, Salt Lake, Kolkata - 700106, India}

\author{S. Ramasesha}
\email{ramasesh@sscu.iisc.ernet.in}
\affiliation{Solid State and Structural Chemistry Unit, Indian Institute of Science, Bangalore 560012, India}

\date{\today}

\begin{abstract}
	Magnetization plateaus are some of the most striking manifestations
	of frustration in low-dimensional spin systems. We present numerical
	studies of magnetization plateaus in the fascinating spin-1/2 skewed
	ladder system obtained by alternately fusing five- and seven-membered
	rings. This system exhibits three significant plateaus at $m = 1/4$, 
	1/2 and 3/4, consistent with the Oshikawa-Yamanaka-Affleck condition. 
	Our numerical as well as perturbative analysis shows that
	the ground state can be approximated by three weakly coupled singlet
	dimers and two free spins, in the absence of a magnetic field. 
	With increasing applied magnetic field, the dimers
	progressively become triplets with large energy gaps to excited states,
	giving rise to stable magnetization plateaus. Finite-temperature
	studies show that $m=1/4$ and 1/2 plateaus are robust and survive
	thermal fluctuations while the $m=3/4$ plateau shrinks rapidly due to
	thermal noise. The cusps at the ends of a plateau follow the 
	algebraic square-root dependence on $B$.
\end{abstract}

\maketitle

\section{\label{sec:intro}Introduction}
\begin{figure}[b]
	\includegraphics[width=3.4 in]{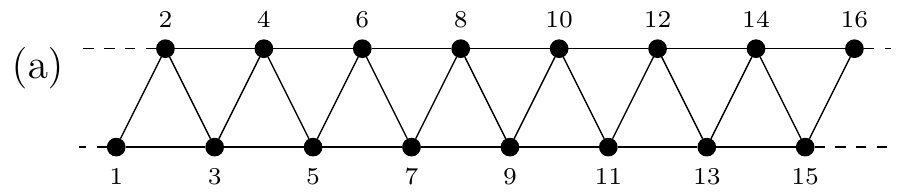} \\
	\includegraphics[width=3.4 in]{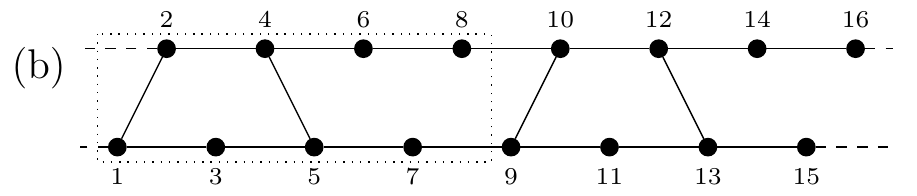} 
	\caption{\label{fig:ladders}(a) The regular zigzag chain and 
	(b) the 5/7 skewed ladder lattice. A unit cell consisting of 
	eight spins is shown in the box with dotted edges.}
\end{figure}

The study of quantum phase transitions in frustrated low dimensional magnets 
has been an active area of research in the last few decades. The simplest 
model of a frustrated magnet is the one dimensional (1D) $J_1-J_2$ model 
where each spin is interacting with its nearest and next nearest neighbors 
with exchange parameters of strength $J_1$ and $J_2$ respectively.~\cite{ckm69a,*ckm69b,hamada88,chubukov91,chitra95,white96,itoi2001,mahdavifar2008,sirker2010,mk2015,soos-jpcm-2016,mk_bow,chubukov91,mk2012,mk2015,vekua2007,hikihara2008,sudan2009,dmitriev2008,meisner2006,meisner2007,meisner2009,aslam_magnon,kecke2007}
This system can be mapped on to a zigzag ladder if odd and 
even sites of the chain are arranged on the two legs of the ladder, as shown 
in Fig.~\ref{fig:ladders}(a).  The frustration in these magnetic systems can give 
rise to many exotic phases in the ground state (gs), such as gapless spin 
liquid,~\cite{hamada88,white96,itoi2001,mahdavifar2008,mk2015,soos-jpcm-2016}
dimers~\cite{ckm69a,*ckm69b,hamada88,chubukov91,chitra95,white96,sirker2010,mahdavifar2008,mk2015,soos-jpcm-2016,itoi2001,mk_bow}
and spiral phases,~\cite{chubukov91,white96,soos-jpcm-2016,sirker2010,mahdavifar2008,mk2012} in the absence 
of an applied magnetic field $B$. These systems become a veritable zoo of phases 
in the presence of a finite magnetic field 
$B$;~\cite{vekua2007,hikihara2008,sudan2009,sirker2010,dmitriev2008,meisner2006,meisner2007,meisner2009}
for example a system with exchange interactions $J_1$ (ferromagnetic) and $J_2$ (antiferromagnetic) 
shows multi-magnon condensation,~\cite{aslam_magnon,sudan2009,kecke2007,hikihara2008,aslam_vchiral}
a vector chiral phase~\cite{chubukov91, hikihara2008,sudan2009,aslam_vchiral} and magnetization 
plateaus.~\cite{meisner2007,tandon99,okunishi_prb2003,*okunishi_jpsj2003} In addition 
to a magnetization plateau phase, frustrated systems can also show kinks and jumps 
in magnetization.~\cite{meisner2007}

There have been sevaral recent attempts to gain a theoretical understanding of the magnetization process in quantum spin chains 
and ladders.~\cite{oya97,cabra97,totsuka98,sakai98,sakai99,honecker2000,okamoto2001,tandon99,okunishi_prb2003,*okunishi_jpsj2003,hida2005,meisner2007,alicea2007}
The magnetization plateau indicates the existence of energy gaps between two consecutive 
magnetic spin sectors in the thermodynamic limit, and one such example is the 
integer spin Heisenberg antiferromagnetic (HAF) chain.~\cite{haldane83a,haldane83b,affleck86}  The energy gap between the gs ($S=0$) and 
the next magnetic excited state ($S=1$) is finite for an integer spin chain with a periodic 
boundary condition, which is the  well-known Haldane gap.~\cite{haldane83a,haldane83b,affleck86}
Therefore, there is an $M=0$ plateau in the gs, and a finite field $B$ of 
strength  $\frac{E_g}{g \mu_B}$ ($E_g$ is the Haldane gap, $g$ 
is the gyromagnetic ratio and $\mu_B$ is the Bohr magneton) is required to 
spin-polarize the system. In fact, the 
Heisenberg spin-1 chain with single ion anisotropy also shows a plateau at 
$m = M/M_{\mathrm{max}} = 1/2$, where $M$ is the magnetization at the plateau and $M_{\mathrm{max}}$ is 
the saturation magnetization.~\cite{nakano98}  The 1D $J_1-J_2$ model with exchange interactions $J_1$ (ferromagnetic)
and $J_2$ (antiferromagnetic) as well as both antiferromagnetic $J_1$ and $J_2$
shows  plateaus at $m$ values of $0$ and $1/3$ for $|J_2/J_1| > 0.6$.~\cite{tandon99,okunishi_prb2003,*okunishi_jpsj2003,meisner2007}
Magnetization plateau at $m=0$ has also been predicted for the ordinary
two-legged ladder, which has finite spin gap in the gs.~\cite{oya97,cabra97}
In the  2D system, magnetization plateaus in a kagom\'e lattice are predicted to be at  
$m=1/9$, 1/3, 5/9 and 7/9,~\cite{morita2018,honecker2002,Schmidt_2006,nishimoto2013,capponi2013,slal_kagome2018}
whereas  for a triangular lattice, the magnetization plateau appears at $m = 1/3$.~\cite{farnell2009}

Experimentally, the magnetization plateau at $m=1/3$ is found in frustrated spin chains 
such as  Cu$_3$(CO$_3$)$_2$(OH)$_2$,~\cite{kikuchi2005,kikuchi2006,gu2006}  and  the spin-1/2 trimer compound 
Cu$_3$(P$_2$O$_6$OH)$_2$.~\cite{hase2006}  Other compounds showing a
1/3 plateau are Ca$_3$Co$_2$O$_6$,~\cite{zhao2010,maignan2004,hardy2004} Sr$_3$Co$_2$O$_6$,~\cite{wang2011}
Sr$_3$HoCrO$_6$,~\cite{hardy2006} SrCo$_6$O$_{11}$~\cite{ishiwata2005} and CoV$_2$O$_6$,~\cite{yao2012,lenertz2011,he2009}
Frustrated ladder compound NH$_4$CuCl$_3$~\cite{shiramura98} shows  two plateaus at 
$m = 1/4$ and 3/4.  

Oshikawa, Yamanaka, and Affleck (OYA)~\cite{oya97} 
established the necessary condition for the occurrence of plateaus in a 1D spin-$S$ 
system by generalizing the Lieb-Schultz-Mattis (LSM) theorem.~\cite{lsm61,affleck86}
The OYA condition for observing a plateau at $m$ is given by
$ S \, p (1 - m) \in \mathbb{Z} $, where $S$ is the spin of a site, $p$ 
is the number of lattice sites per unit cell, and $\mathbb{Z}$ represents the set of positive integers. 
The condition is further generalized to $n$-leg ladder,~\cite{cabra97,cabra98} which is given as 
$ n \, S \, p \, (1 - m) \in \mathbb{Z}$. The Haldane chain is a special case of 
this condition in which $n = p = 1$, and for integer $S$ a plateau could  occur 
at $m = 0$.

In this paper we are interested in the skewed ladder system which is a variant 
of a zigzag ladder with periodically missing  rung bonds.~\cite{geet}
The 5/7 system, which corresponds to fused azulenes,
shows a high-spin gs for large $J_1/J_2$ in the thermodynamic 
limit. Thomas {\it et al.} showed that fused azulene systems have both a high-spin gs and
permanent electric polarization in the thermodynamic limit.~\cite{thomas2012}
In addition, it is possible to fabricate such ladders as defects 
in graphene layers or at the grain boundaries in 
graphene.~\cite{graphene_defect_nature_2011, arindam_ghosh_2015, balasubramanian2019}
The electrons in the $2p_z$ orbital of the Carbon atoms in these ladders form a 
strongly correlated band that can be modeled by a long-range interacting model 
such as the Pariser-Parr-Pople (PPP) model.~\cite{pariser_parr53,*pople53}
The PPP model transforms into an isotropic spin-1/2 Heisenberg model with 
antiferromagnetic exchange interactions. The magnetic properties of these two 
models are closely related.~\cite{thomas2012}
The behavior of these systems in the presence of a static axial magnetic field 
had not been studied so far. Here, we analyze the behavior of this skewed ladder in the 
presence of an external magnetic field. 

The skewed ladders can be of various 
types,~\cite{geet} but in this paper we study only the 5/7 skewed ladder shown 
in Fig.~\ref{fig:ladders}(b). In this system there are eight spins and  ten bonds 
per unit cell. There are two bonds with exchange interaction $J_1$ and eight 
bonds with $J_2$.   This system shows a high-spin gs in the large $J_1$ ($> 2.35 J_2$) 
limit, and one-quarter of the spins in each unit cell are connected through the effective 
ferromagnetic interaction while the remaining six spins form three singlet dimers.~\cite{geet}
The OYA condition predicts the 
possibility of plateaus at $m = 0$, 1/4, 1/2 and 3/4 for this system. In this paper, 
we show that there are indeed three plateaus at $m=1/4$, 1/2 and $3/4$ in the 5/7 
skewed ladder system; in the regular zigzag chain a lone plateau at
$m=1/3$ is observed, although the OYA condition predicts two more plateaus at 
$m=0$ and 2/3.~\cite{tandon99,okunishi_prb2003,*okunishi_jpsj2003,meisner2007}
These plateaus in the 5/7 ladder are formed because of the strong dimer 
formations in the system. The existence of four plateaus in a ladder is not found 
in the literature.  We also explore the cusp at the ends of the plateaus.

This paper is divided into four sections. In section~\ref{sec:model} we discuss the model 
Hamiltonian  and the numerical method. The results are presented in section~\ref{sec:results} which 
has three subsections. The discussion of results are presented in section~\ref{sec:sum}.

\section{\label{sec:model}Model and Method}
In Fig.~\ref{fig:ladders}(b) we show schematically a 5/7 ladder. 
All exchange interactions between the spins are antiferromagnetic in nature. 
The sites are numbered such that odd-numbered sites 
are on the bottom leg and even-numbered sites are on the top leg. In this scheme, the rung bonds 
are the nearest-neighbor exchanges $J_1$ and the bonds on the legs are the
next-nearest-neighbor exchanges. The nearest-neighbor exchange $J_1$ is taken in units 
of $J_2$, which defines the energy scale. In the presence of an axial magnetic field 
$B$, the model Hamiltonian can be written as
\begin{eqnarray}
	H_{5/7} &=& J_1 \sum_{i=0}^n \left(\vec{S}_{8i+1} \cdot \vec{S}_{8i+2} + 
	\vec{S}_{8i+4} \cdot \vec{S}_{8i+5} \right) + \nonumber \\
	& & J_2 \sum_{i=0}^n \sum_{k=1}^8
	\vec{S}_{8i+k} \cdot \vec{S}_{8i+k+2} - B \sum_{i=0}^n \sum_{k=1}^{8} S^z_{8i+k}
	\label{eq:ham}
\end{eqnarray}
where $i$ labels the unit cell and $k$ is the spin in the unit cell, as shown in 
Fig.~\ref{fig:ladders}(b). The first and second terms denote the rung exchange  interaction $J_1$ and 
the interaction along the legs $J_2$, and the third term of the Hamiltonian gives 
the interaction of the spins with an axial field $B$ in units of $J_2 / g \mu_B$.

The Hamiltonian in Eq.~\ref{eq:ham} is many-body in nature, therefore we need 
to deal with a large number of degrees of freedom. We use the density matrix 
renormalization group (DMRG) method to handle the large degrees of 
freedom.~\cite{white-prl92,white-prb93,mk2010}
The DMRG method is based on systematic truncation of  irrelevant  degrees 
of freedom. The dimension of the chosen effective density matrix $m$ (which is 
also the number of block states, not to be confused with fractional magnetization) 
varies by up to 400, and the 
truncation error of the density matrix is less than $10^{-11}$. We also carry out 
five to six  finite DMRG sweeps for satisfactory convergence of the eigenstate. 
The growth sequence of the system  is the same as that in our earlier work.~\cite{geet}
The lowest eigen states  in all $M_S$ sectors, from zero to $N/2$, are calculated to 
find the total spin $S_{gs}$ of the gs from the condition $E_0(S_{gs} - 1) = 
E_0(S_{gs}) < E_0(S_{gs} + 1)$. An exact diagonalization (ED) technique is used 
to calculate the gs properties of small systems.

\section{\label{sec:results}Results and Discussions}
The 5/7 ladder system has eight spins per unit cell,  and  in the large-$J_1$ 
limit, the ground state can be represented as six spins 
forming three singlet dimers and  weak ferromagnetic interaction between the 
remaining two spins on the same legs in the two adjacent rings. Therefore, in the thermodynamic 
limit, for $J_1/J_2 > 2.35$, the magnetization in the gs is $m=1/4$.  
In the strong $J_1 (> 2.35)$ limit, the gs has $m=1/4$ for $B=0$, even for small systems.  
In this paper we  focus on the gs properties 
in the presence of an axial magnetic field, and we show that there are three
plateau phases at $m=1/4$, 1/2 and 3/4 as predicted by the generalized 
Lieb-Schultz-Mattis theorem.~\cite{lsm61,affleck86,oya97} To understand the plateau phases, we 
analyze the gs energies, spin densities, and spin-spin correlations in the presence 
of the applied magnetic field, $B$. The Hamiltonian in Eq.~\ref{eq:ham} conserves
$M_S$, therefore field dependent gs energies are obtained simply by adding the 
Zeeman term to the zero field energies,
\begin{equation}
	E(M_S, B) = E_0(M_S, B=0) -  B M_S,
	\label{eq:nrgB}
\end{equation}
where $M_S$  is  the $z$-component of the total spin. 
The change in magnetization from $M_S$ to $M'_S$ occurs at the crossing of the 
$E(M_S, B)$ and $E(M'_S, B)$ lines.

The gs of this  model was studied in the absence of a magnetic field  in Ref.~\onlinecite{geet}, 
and it shows a  high spin ferrimagnetic gs even at $J_1 = 1$ and a re-entrant 
antiferromagnetic phase for $1.75 < J_1 < 2.18$. The gs has $S = n$ 
(the number of unit cells) for $J_1 > 2.35$. In  this phase the spins at sites 
3, 7, 11, \ldots have effective  ferromagnetic 
interactions.~\cite{geet} The spin density $\rho_i$ at site 
$i$ and  axial spin correlation $C(r)$ between spins at sites $i$ and $i+r$  
are defined as
\begin{equation}
	\label{eq:rho_def}
	\begin{aligned}
		\rho_i &= \langle gs \mid S_i^z \mid gs \rangle \\
		C(r) &= \langle gs \mid S_i^z S_{i+r}^z \mid gs \rangle.
	\end{aligned}
\end{equation}

The results are presented in the following four subsections. First the $m-B$ 
curves and their corresponding energies are presented in~\ref{subsec:A}. 
To understand all three  plateau phases, the  spin density $\rho_i$ at site `i'
and the spin-spin correlation function $C(r)$ (Eqn.~\ref{eq:rho_def}) are studied
for different $M_S$ values in ~\ref{subsec:B}. From these we can infer the gs 
spin configurations at different magnetization plateaus. Here we
also study the cusp singularities and thermal stability of plateaus.
The analytical perturbation theory to support this is discussed 
in~\ref{subsec:C}. 

\subsection{\label{subsec:A}Plateau phases in 5/7 ladder}
\begin{figure}
	\includegraphics[width=3.4 in]{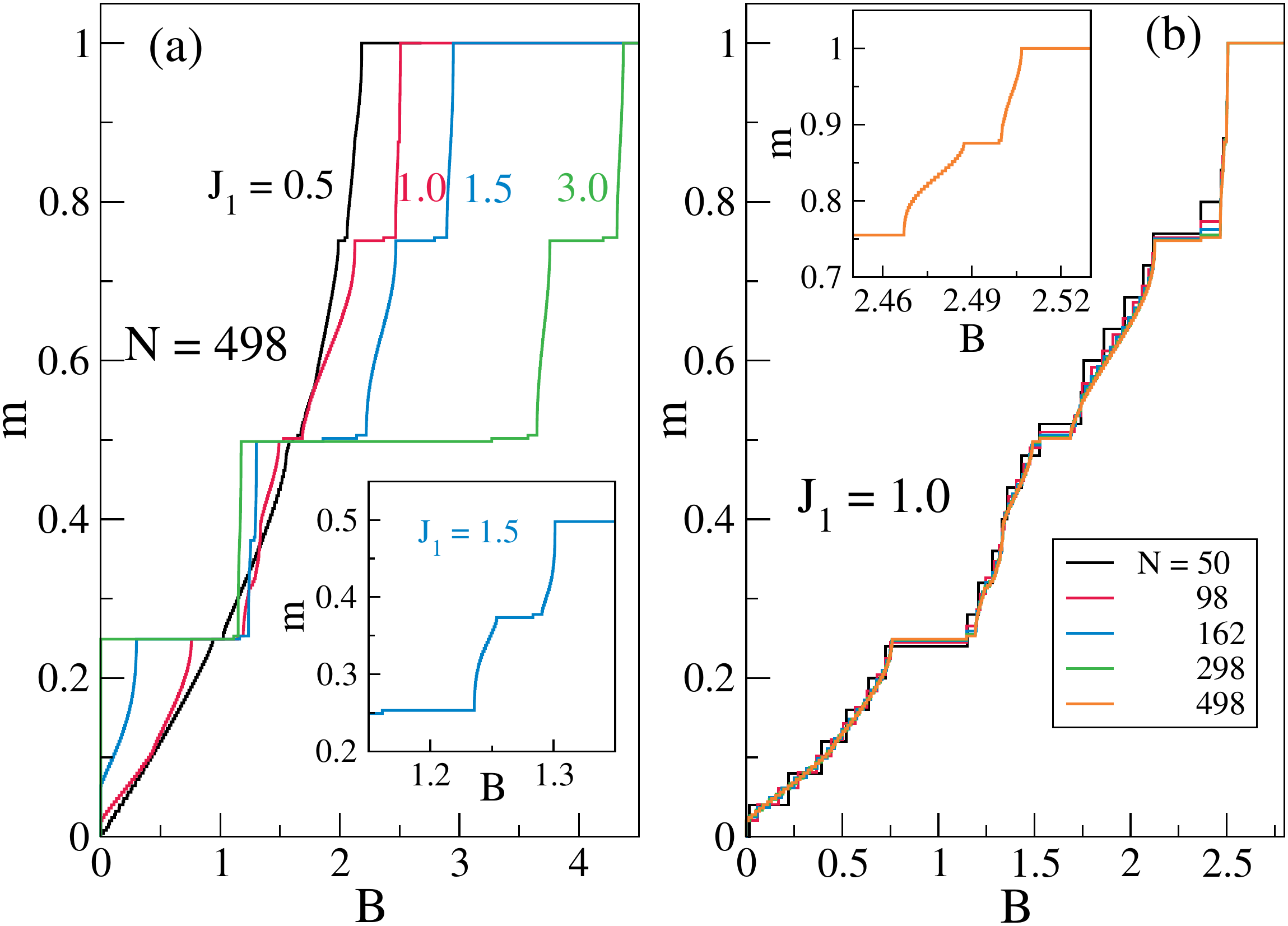}
	\caption{\label{fig:mh}(a) $m-B$ curve for 5/7 skewed ladder for $J_1 = 0.5$,
	$1.0$, $1.5$ and $3.0$ for $N=498$ sites. Inset shows an $m=3/8$ plateau for 
	$J_1 = 1.5$. (b) The finite size effect of the $m-B$ curve with $J_1 = 1.0$ 
	for five system sizes. Inset shows an $m=7/8$ plateau at this $J_1$ value.}
\end{figure}
\begin{figure}
	\includegraphics[width=3.4 in]{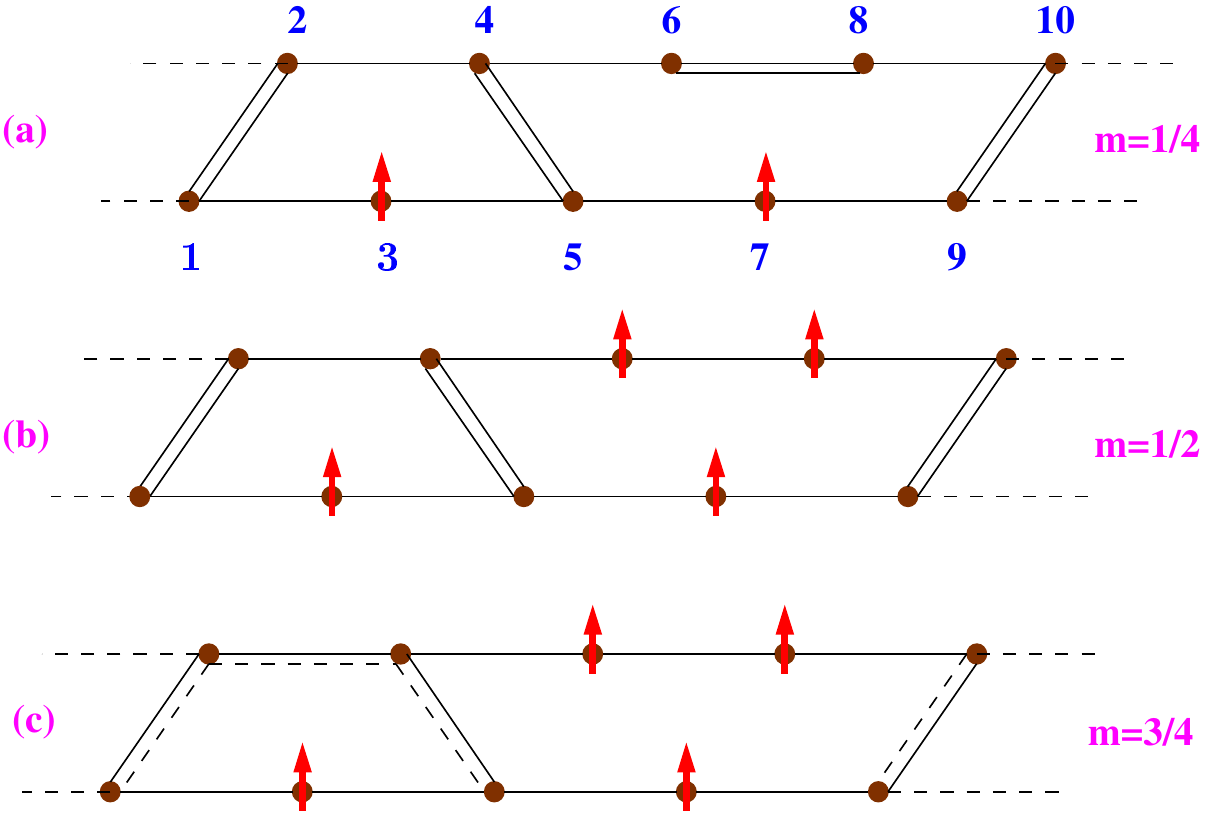}
	\caption{\label{fig:schematic}Schematic representation of gs in terms
	of singlet dimers and aligned free spins at (a) $m=1/4$ plateau, (b) $m=1/2$ 
	plateau and (c) $m = 3/4$ plateau, where broken lines imply a triplet
	delocalized over four sites.}
\end{figure}
We plot the $m-B$ curve for different strengths of the rung exchange $J_1$ in 
Fig.~\ref{fig:mh} for $N=498$ spins (62 unit cells), and we note that the 
m=1/4 and 3/4 plateaus appear for  $J_1 \geq 0.4$, whereas,  the other plateau
at $m=1/2$  appears for $J_1 \geq 0.9$. Let us define $B_{i}^L$ and $B_i^U$ as 
the lower and upper critical values of the magnetic field for the $i^{\mathrm{th}}$ 
plateau where $i = 1$, 2, 3 for the three plateaus at $m = 1/4$, 1/2 and 3/4 respectively. 
We note that onset field of the first plateau at $m=1/4$ decreases as $J_1$ 
increases, while the width of the plateau $w_1 = (B_1^U - B_1^L)$ increases with 
increasing $J_1$. The width of the $1/4$ plateau increases linearly with
$J_2$ in the large $J_1$ limit, as shown analytically in 
subsection~\ref{subsec:C}.  We notice that a finite field is required to 
reach $m=1/4$ for small $J_1$; in this limit of $J_1$, the rung dimers are weak 
and a finite field is required to align spins 3 and $7$ in the field direction. 
In the  $J_1 >2.35$ limit, all the rung singlets are strong, and an effective 
ferromagnetic interaction develops between spin $3$ and $7$ as shown in 
Fig.~\ref{fig:schematic}(a). Thus the field required to attain the 1/4 plateau decreases to zero.
Along both legs, most bonds are weak except the generic 6-8 bond in each unit cell. 
Therefore, the next plateau occurs when the 6-8 bond breaks and the system enters 
another locked phase with $m=1/2$ (Fig.~\ref{fig:schematic}(b)). In this case, most of the magnetic contribution 
comes from the ferromagnetically aligned spins at sites 3, 6, 7, and $8$ in the unit 
cell (Fig.~\ref{fig:schematic}(b)). On further ramping the field $B$, the singlet involving
sites 1, 2, 4 and 5 flips to yield a triplet, and the system locks into the $m=3/4$ 
plateau phase (Fig.~\ref{fig:schematic}(c)). At fields greater than the saturation field $B_{\mathrm{sat}}$,
all singlet bonds in every unit cell are broken, and the system goes to a completely 
polarized state.  
\begin{figure}
	\includegraphics[width=3.4 in]{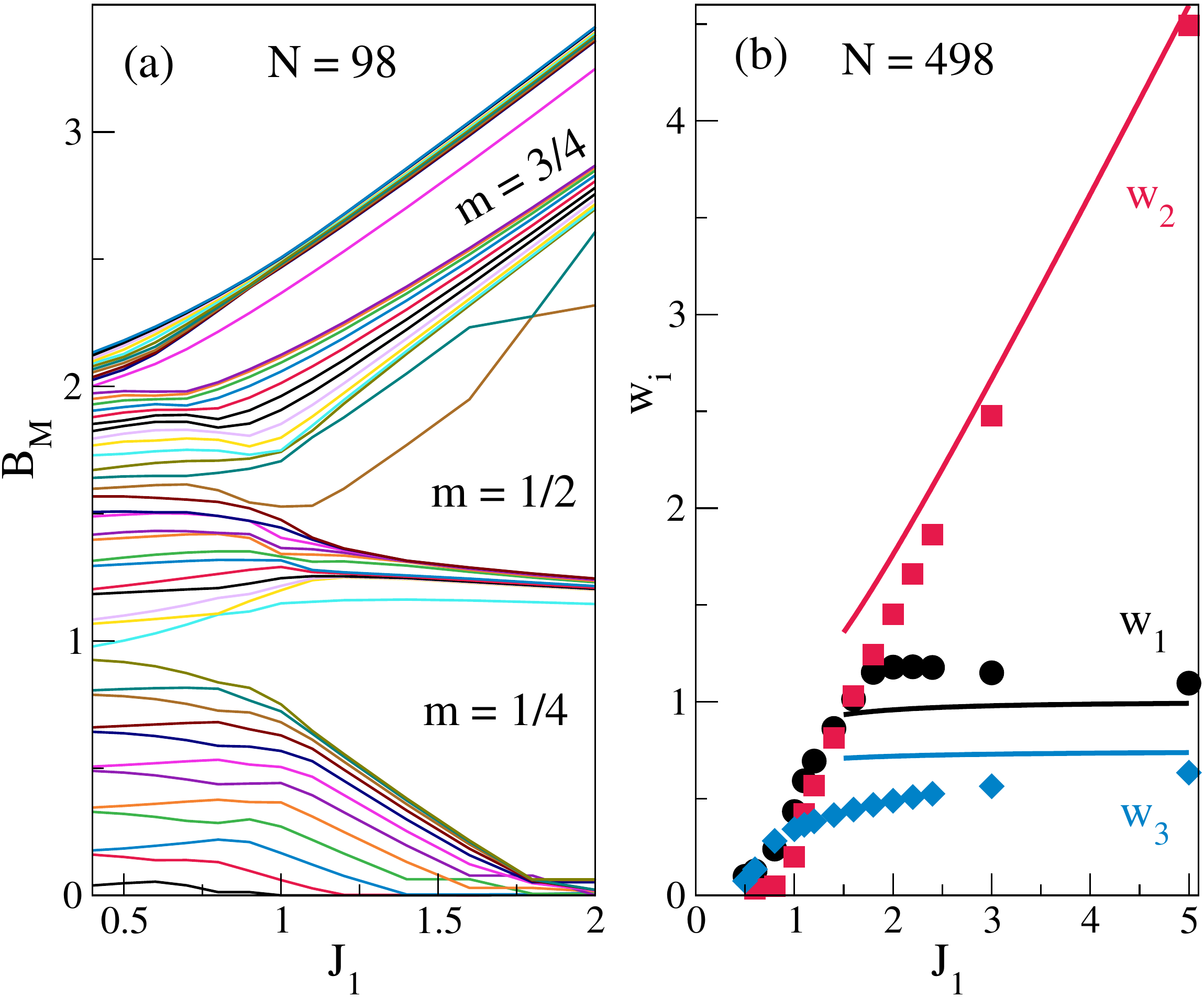}
	\caption{\label{fig:width} (a) The magnetic field $B_M$ required to 
	close the energy gap between successive lowest energy $M_S$ states vs the strength
	of nearest neighbor exchange $J_1$. $B_M$ is in units of $J_2 / g \mu_B$.
	The $M$ values increase from $M=1$ for the bottom curve to $M=49$ 
	for the uppermost curve. At large $J_1$, these curves collapse into four
	groups with $m = M/M_{\mathrm{max}} = 1/4$, $1/2$, $3/4$ and $1$.
	(b) The width of the plateaus $w_1$, $w_2$ and $w_3$ for $m=1/4$, 
	$1/2$ and $3/4$ plotted as a function of $J_1$ for a system with
	$N=498$ spins. The solid lines are the plateau widths calculated 
	from the perturbation theory in the large $J_1$ limit (see~\ref{subsec:C}).}
\end{figure}

Other than these three significant plateaus, there is a narrow plateau at $m=3/8$ for
$1.2 < J_1 < 2.5$ and another at $m=7/8$ for $J_1 = 1.0$, as shown in the insets of 
Figs.~\ref{fig:mh}(a) and (b) respectively. These plateaus disappear at larger 
$J_1$ values. In the OYA condition, if we consider an enlarged unit cell of
$p = 16$ sites, then there is the possibility of plateaus at $m=1/8$, $3/8$, $5/8$
and $7/8$ other than the plateaus mentioned earlier. However, we only find 
two of them namely, at $m=3/8$ and $7/8$ for a particular range of $J_1$ values.
Again we can conjecture that the $m = 3/8$ plateau appears 
because at the corresponding magnetic field one of the singlet bonds 6-8 or 14-16 
(Fig.~\ref{fig:ladders}b) in the enlarged unit cell breaks. Similarly, the 
$m = 7/8$ plateau could arise when one of the triplets involving sites 
[1,2,4,5] or [9,10,12,13] leads to four unpaired spins. At large $J_1 > 2.5$, 
these small plateaus disappear, perhaps because the correlations become short-ranged 
and the coupling between unit cells becomes weak.

To estimate the width of plateaus we plot the magnetic field, $B$, required 
to achieve successively higher $M_S$ states for a given $J_1$.   Here, we 
define magnetic field  $B_M$ as the magnetic field required to close the gap 
between the $M_S = M$ and $M_S = M+1$ state i.e., 
\begin{equation}
	B_M = \frac{E_0(M+1) - E_0(M)}{g \mu_B}.
\end{equation}
In Fig.~\ref{fig:width}(a) we notice that the $B_M$ curves collapse into 
four bands for large $J_1$: the first band corresponds to $m=1/4$, the second 
band corresponds to the $m=1/2$, the third corresponds to $m=3/4$, and the fourth
corresponds to $m=1$. This is a consequence of the fact that for large $J_1$, the 
ground state has a spin $p/8$, and excited states with a finite gap in the thermodynamic
limit have spins $p/4$, $3p/8$ and $p/2$
where $p$ is the number of spins in a unit cell. There are three plateaus of 
significance corresponding to $m=1/4$, 1/2 and 3/4. The width of the 1/4 plateau at
large $J_1$ corresponds to the magnetic field difference between the $m = 1/4$ and 
$m=1/2$ bands. This remains independent of $J_1$ at large $J_1$. 
The magnetic field $B_M$ for $m=1/2$ is independent of $J_1$
while that of $B_M$ with $m=3/4$ increases linearly with $J_1$ at large $J_1$. Therefore in this limit, the
plateau width for $m=1/2$ increase linearly with $J_1$. The width of the $m=3/4$ plateau
is independent of $J_1$ as the $B_M$ for $m=3/4$ and $m=1$ both increase linearly with
$J_1$ and their difference is independent of $J_1$, for large $J_1$. This can be seen
in Fig.~\ref{fig:width}(b), where the plateau width for magnetization calculation is
plotted as a function of $J_1$.

\begin{figure}
	\includegraphics[width=3.4 in]{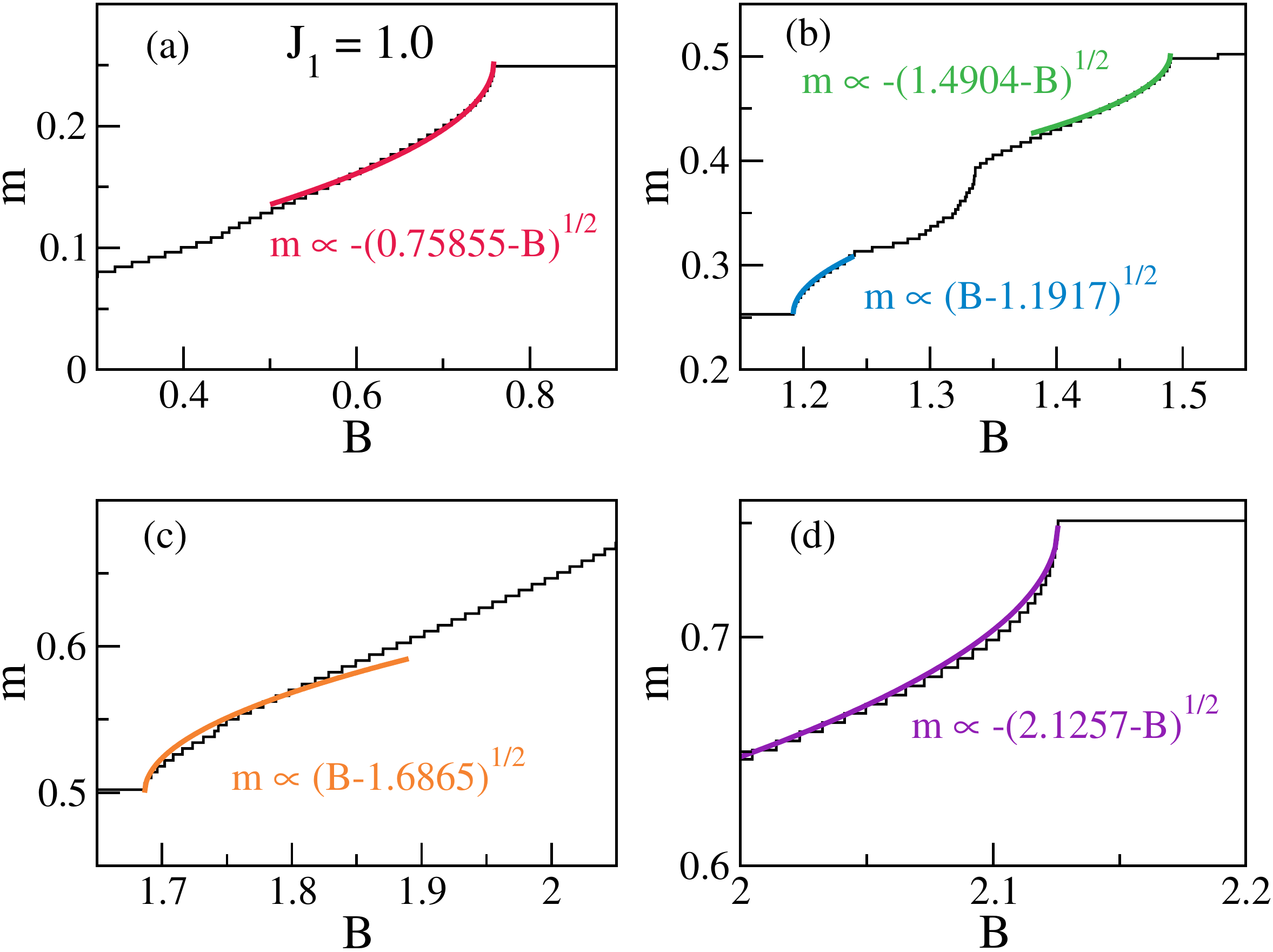}
	\caption{\label{fig:fitting} Behavior of magnetization at the plateau ends 
	for 5/7 ladder at $J_1 = J_2 = 1.0$ for $N=498$ spins (62 unit cells)
	(a)$m = 1/4$ plateau at $B_c^L$, (b) $m = 1/4$ plateau at $B_c^U$, 
	(c) $m = 1/2$ plateau at $B_c^U$ and (d) 3/4 plateau at $B_c^L$. The magnetization 
	near the plateaus obey $\propto (B-B_c)^{1/2}$ and the numerical $m$ vs $B$ 
	dependence at the plateau ends are shown in the figure.}
\end{figure}

At the ends of the plateaus, the magnetization should vary as $m(B) - m(B_{c}) \propto 
(B - B_{c})^{1/2}$,~\cite{bonner_fisher} where $B_c$ is the magnetic field at the end of the plateau under consideration. 
In fact for 5/7 ladder, there are at least four cusp singularities (at $m=1/4$, 1/2 and  3/4)
in the $m-B$ curve for the $J_1 = 1.0$ case with $N = 498$ spins (62 unit cells) 
as shown in Fig.~\ref{fig:fitting}.  The magnetization curves 
near the plateau are fitted with $m \propto (B-B_c)^{1/2}$. The $B_c^L$ ($B_c^U$) 
for this system for $J_1=1$ are $0.759$ ($1.192$), $1.490$ ($1.687$), and $2.126$ for $m=1/4$, 
1/2 and 3/4 respectively(Fig.~\ref{fig:fitting}). At $J_1 = 1.0$ in Fig.~\ref{fig:fitting}(b), 
we also see a cusp like behavior for $m \approx 0.31$ and $\approx 0.39$. This corresponds
to viewing the spin system as belonging to a larger unit cell, as near those values the onset of 
new plateaus occurs, assuming a larger magnetic unit cell.

\begin{figure}
	\includegraphics[width=3.4 in]{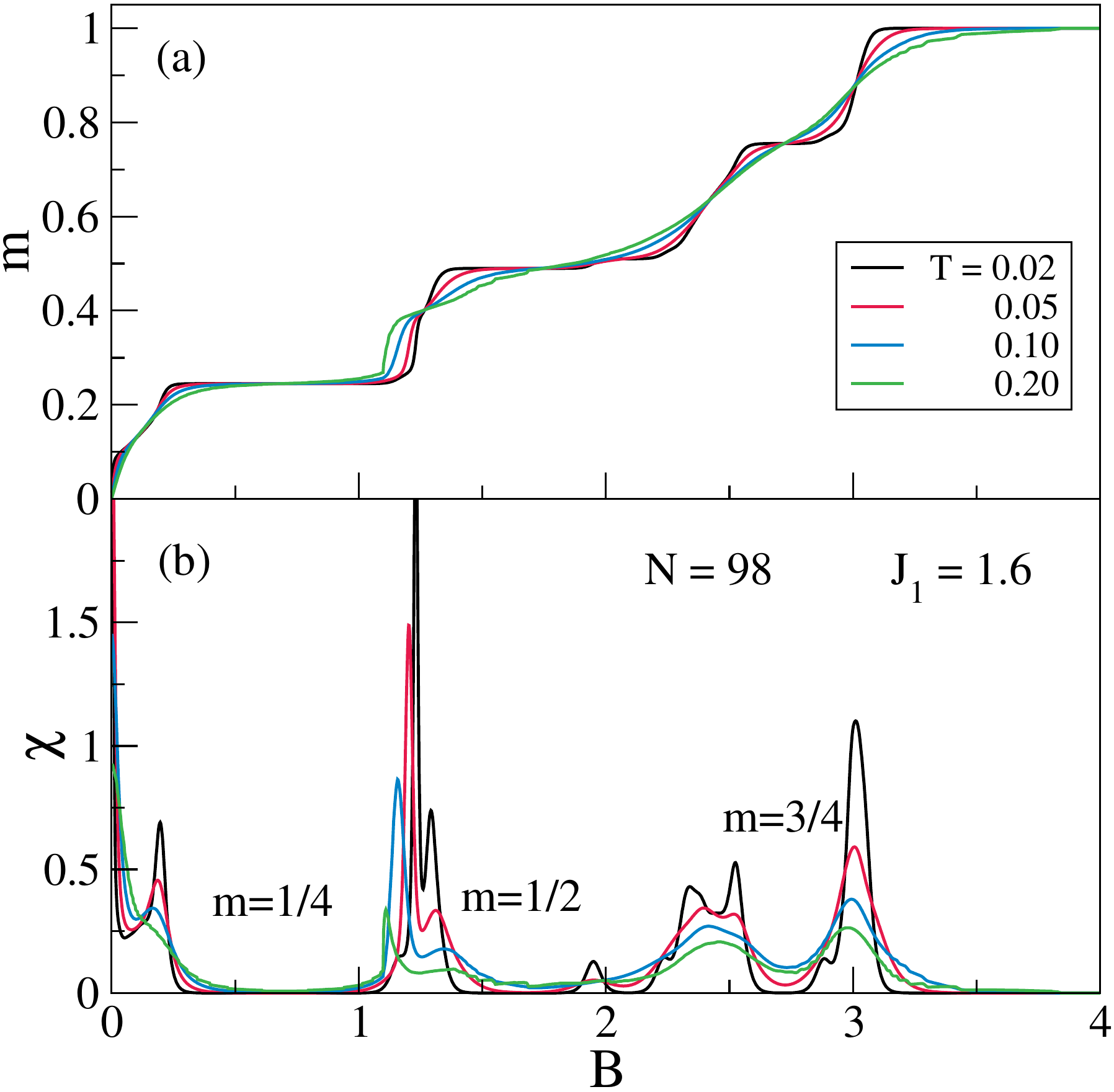}
	\caption{\label{fig:magtemp}$m-B$ curve at four temperatures
	$T/J_2 = 0.02$, 0.05, 0.10 and 0.20 for $J_1 = 1.60$ and $N = 98$ 
	spins.}
\end{figure}
For sufficiently large $J_1$,  the plateaus are wide and the energy gap
near each plateau is large. Any practical application of these plateaus 
are feasible only if they are stable to thermal fluctuations.  The finite 
temperature behavior of the magnetization plateau of the system is shown in 
Fig.~\ref{fig:magtemp} (a)  with $J_1=1.6$ for $N = 98$ spins. We have 
used a hybrid ED-DMRG method with the average density matrix taken over many states in each
$M_S$ sector.~\cite{hybrid} We have used $300$ energy eigenvalues from each $M_S$ 
sector for thermal averaging of the magnetization. 
Since we are focusing on the low-temperature properties 
of the system, retaining 300 low lying states in each $M_S$ sector should
be accurate as higher excited states are practically 
inaccessible at low temperatures. As shown in Fig.~\ref{fig:magtemp} (a), 
the plateaus at $m=1/4$ and 1/2  are robust at finite temperatures, while 
the plateau at 3/4 survives only up to $T/J_2 = 0.05$. The derivative  
$\chi = \frac{\mathrm{d}m}{\mathrm{d}B}$ vs $B$ is shown in 
Fig.~\ref{fig:magtemp} (b). In the plateau regime the susceptibility $\chi$
vanishes at $T=0$ whereas, it is finite when the plateaus are perturbed by  the 
thermal fluctuations.  We notice that 3/4 plateau is very much susceptible to the thermal 
fluctuation as $\chi$ is finite even at low temperatures, however 
$m=1/4$ and 1/2 plateaus survive thermal fluctuations. 
This can be attributed to small energy gaps, between the successive lowest $M_S$ states, 
near the $m=3/4$ plateau, which leads to a gradual change in  magnetization at low 
temperatures. We also note that the plateau width is smallest for the $m=3/4$ 
plateau at 0 K, which also implies that the plateau should disappear on 
slight warming. 

\subsection{\label{subsec:B}Spin density and correlation function in the plateau phase}
\begin{figure}
	\includegraphics[width=3.4 in]{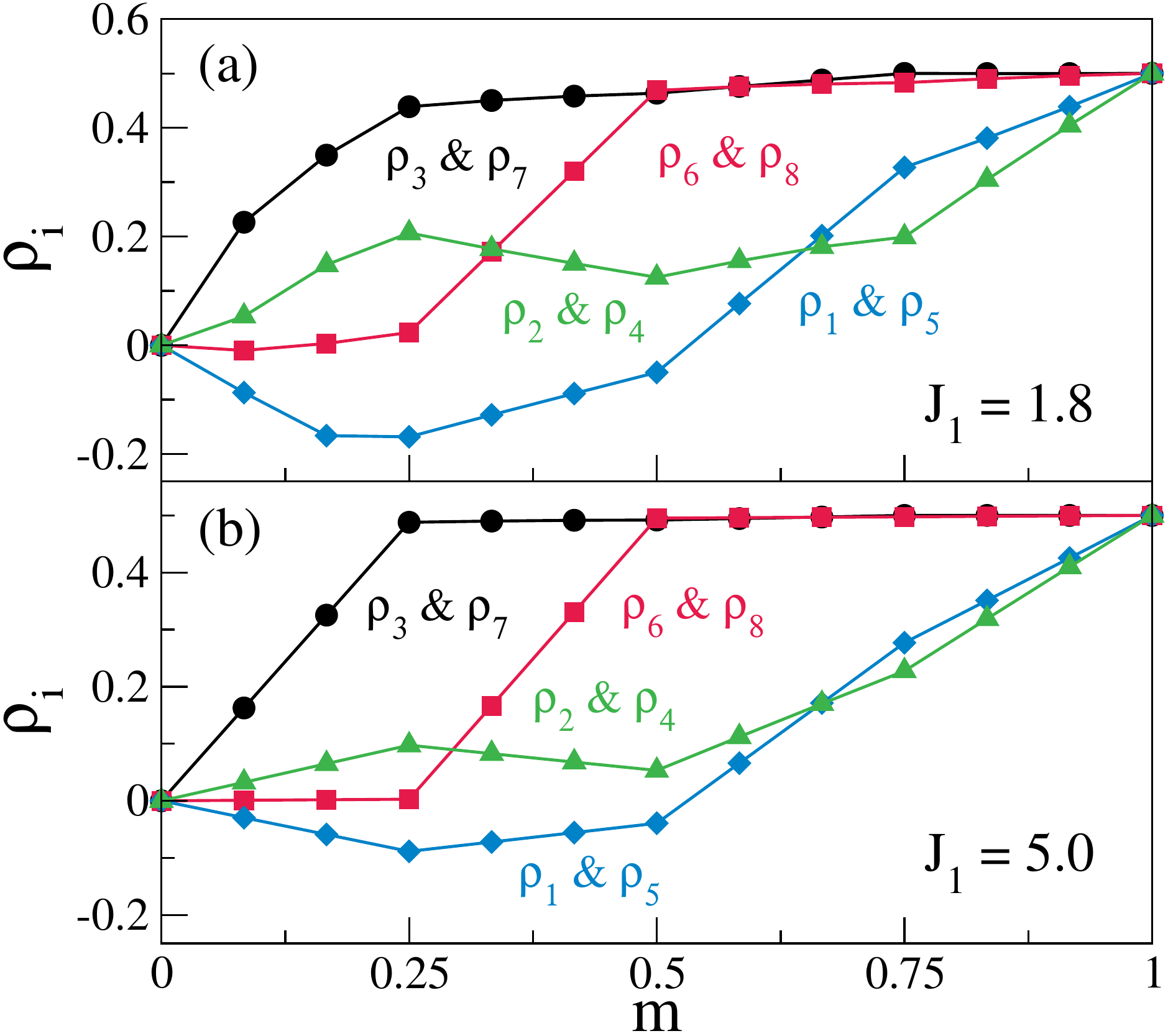} 
	\caption{\label{fig:spden}The spin densities at different sites in a
	unit cell of 5/7 ladder of $N=24$ sites with periodic boundary condition.}
\end{figure}

To identify the spins that are aligned along the applied field in a given plateau,
we study the spin densities and spin-spin correlation functions with 24 spins
under a periodic boundary condition (PBC). For a system of 24 spins, $M_S$ varies
from 0 to 12. We obtain spin densities at all the sites for the lowest eigenstate
in each positive $M_S$ sector. Based on the calculated spin densities we find that 
sites (1, 5), (2, 4), (3, 7) and (6, 8) have very nearly the same spin densities in 
all the $M_S$ sectors of the system. We show in Fig.~\ref{fig:spden} the variation
of spin densities with $m = M_S / 12$ for $J_1 =1.8$ and $5.0$  corresponding to below and 
above the critical rung interaction $J_{1c} =2.35$.

The spin densities at site numbers 3 and 7 increase  quickly with $m$, which
attains the value  $\rho_3 \simeq \rho_7 \sim 0.5$ at $m = 1/4$ for both $J_1 =1.8$ and $5.0$. 
In fact for large value of $J_1 >2.35$,   $\rho_3$ and $\rho_7$ are both $0.5$ 
without any field as $m=1/4$ is the gs.   The spin densities at 6 and 8 ($\rho_6$ and $\rho_8$)
increases linearly with $m$ between $m=1/4$ and 1/2, and these go to 0.5 for $M_S = 6$.  
At the  magnetic field at which this state becomes the gs, the 6-8 singlet bond breaks and 
becomes a triplet bond. The spins at the remaining sites, namely 1, 2, 4, and 5 
form a singlet and have very low spin densities. This singlet state 
transitions to a triplet for the $m=3/4$ plateau and the gs has $M_S = 9$. In this
state the singlet formed by the spins 1, 2, 4 and 5 is canted to a triplet while
the spins at other sites are ferromagnetically aligned. Hence, the spin densities
$\rho_1$, $\rho_2$, $\rho_4$ and $\rho_5$ are almost equal beyond $m=0.75$.

\begin{figure}
	\includegraphics[width=3.4 in]{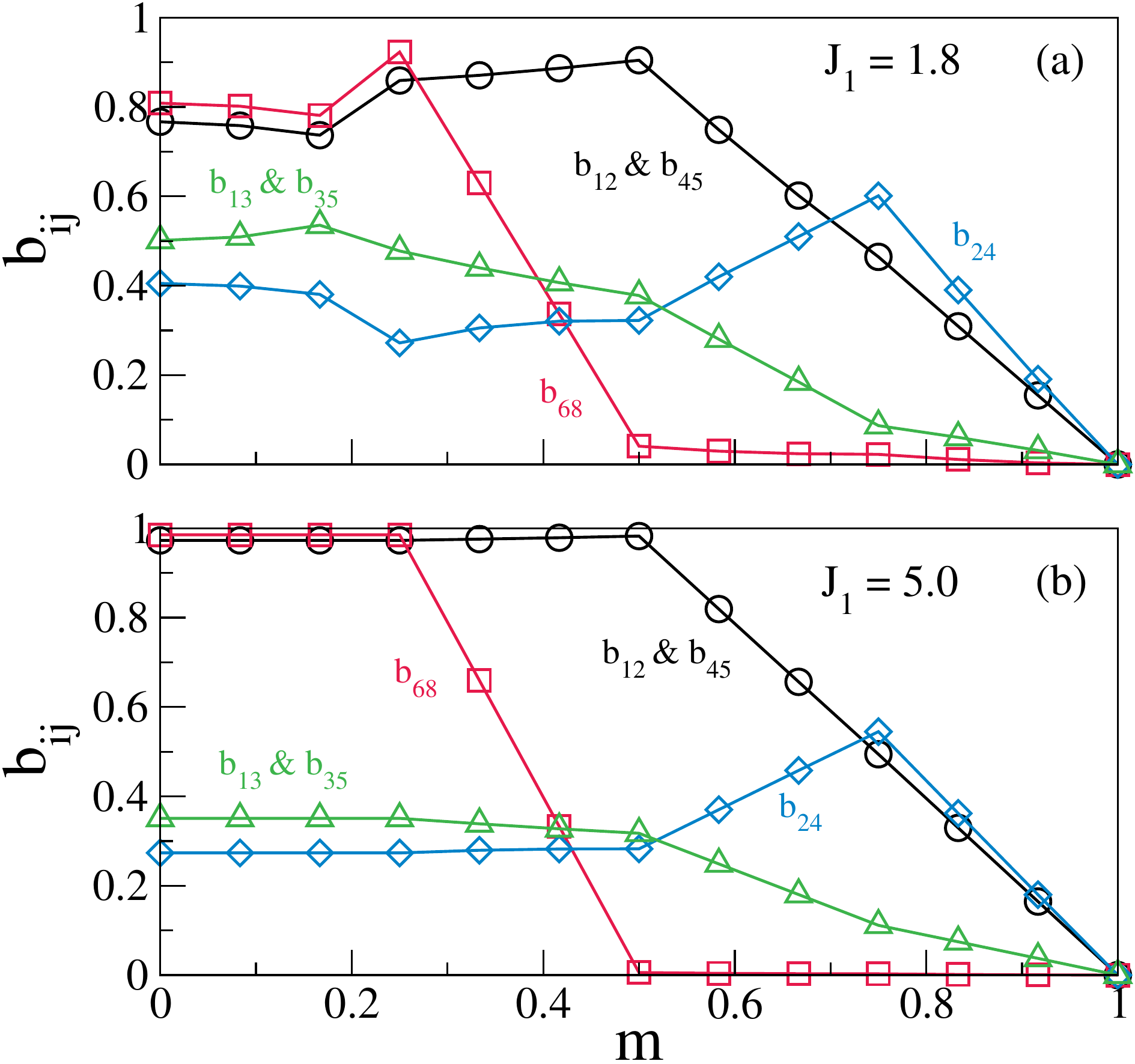}
	\caption{\label{fig:bndord}The bond energies or the bond order for different bonds in a
        unit cell of 5/7 ladder of $N=24$ sites with periodic boundary condition.}
\end{figure}
To understand the effective interactions between the neighboring spins we study 
the bond  energies or magnetic bond order $b_{i,j}=- \langle \psi_{gs} \mid\vec{S}_i \cdot \vec{S}_{j} - \frac{1}{4}
\mid \psi_{gs} \rangle $ where the sites $i$ and $j$ are connected either by a $J_1$ or a $J_2$ interaction. 
There are ten such bonds in a unit cell, and we show only dominant bond-orders in a unit 
cell in Fig.~\ref{fig:bndord}. Only the bonds (1, 2), (4, 5), (6, 8), and (2, 4) 
are significant, and others are small and increase with $m$. 
We find $b_{12} = b_{45}$ and $b_{13} = b_{35}$ for all $M_S$ values. For degenerate
states we obtained $b_{ij}$ by diagonalizing the bond-order matrix in the basis
of degenerate states. In Fig.~\ref{fig:bndord}, we show the variation of $b_{ij}$
with $m$. In the large-$J_1$ case, we note that for $m=1/4$, the (6, 8), (1, 2)
and (4, 5) bonds are singlets. The (1, 3), (3, 5) and (2, 4) bonds are weak and have
a large triplet component. The (3, 7) bond (not shown in the figure) always remain
triplet. When $m$ reach the value of $0.5$, the (6, 8) singlet breaks and a triplet
is obtained. This is true even for small $J_1$ (Fig.~\ref{fig:bndord}a). As the $m$
value increases further, (1, 2) and (4, 5) bonds resonate between a singlet and a 
triplet, but the (2, 4) bond becomes a singlet. This can be represented as a state
in which resonance involves the valence bond pairing $[1,2](4,5) \leftrightarrow [2,4](1,5)
\leftrightarrow (1,2)[4,5]$ where square brackets imply singlet pairing and parentiheses
imply triplet pairing of spins at sites indicated inside the parentheses. All the
singlet bonds are broken when $m=1$ as is to be expected.

\subsection{\label{subsec:C}Perturbation calculations for plateau phases}
The numerical studies indicate that we can develop an understanding of the
plateau phases from a perturbative approach in the large-$J_1$ limit. Hence,
we treat the $J_2$ terms in the spin Hamiltonian as a perturbation over the 
$J_1$ terms. Within a unit cell the $J_1$ term provides interactions between
spins $S_1$ and $S_2$, and $S_4$ and $S_5$, while the $J_2$ term operates between 
spin pairs $S_2$, $S_4$; $S_4$, $S_6$; $S_6$, $S_8$; $S_1$, $S_3$; $S_3$, $S_5$;
and $S_5$, $S_7$. Since the spins $S_6$ and $S_8$ experience only $J_2$ 
interaction, we include this interaction also in the unperturbed Hamiltonian.
Thus for a perturbation calculation, the zeroth order Hamiltonian over one
unit cell under a periodic boundary condition is given by
\begin{eqnarray}
	H_0 &=&  J_1 \left(\vec{S}_{1} \cdot \vec{S}_{2} + 
	      \vec{S}_{4} \cdot \vec{S}_{5}\right) + J_2 \vec{S}_{6} \cdot \vec{S}_{8} \nonumber \\
	      & & -B \left(S_3^z + S_7^z \right).
\end{eqnarray}
Other terms of the Hamiltonian involve $J_2$ interaction and can be treated 
as perturbation $H_1$ given by
\begin{eqnarray}
	H_1 &=& J_2 \sum_{k=1}^8 \vec{S}_{k} \cdot \vec{S}_{k+2} - 
	J_2 \vec{S}_{6} \cdot \vec{S}_{8} \nonumber \\
        & & \qquad - B \left(S^z_{tot} - S^z_3 - S^z_7 \right),
\end{eqnarray}
where a cyclic boundary condition is implied in the summation.
The ground state of the unperturbed system in terms of the spin couplings 
can be written as 
\begin{eqnarray}
	\mid \psi_0 \rangle &=& \mid \overline{1 \quad 2} \ \uparrow_3 \ 
	\overline{4 \quad 5} \ \widefrown{6 \ \uparrow_7 \ 8} \rangle.
	\label{eq:gsappr}
\end{eqnarray}
In Eq.~\ref{eq:gsappr}, the convention we follow is $ |\overline{i \quad j} \rangle = \frac{1}{\sqrt 2} 
\left\{ \mid \uparrow_{i} \downarrow_{j} \rangle - \mid \downarrow_{i} \uparrow_{j} \rangle \right\}$.
The zeroth order energy of the system can be obtained by operating the zeroth
order Hamiltonian $H_0$ on the state $\mid \psi_0 \rangle$ in Eq.~\ref{eq:gsappr},
and it is given by
\begin{eqnarray}
	\mathcal{E}_0 &=& -\frac{3}{2} J_1 - \frac{3}{4} J_2 - B,
\end{eqnarray}
The first order contribution from $H_1$ vanishes as the matrix elements 
$\langle \psi_0 \mid H_1 \mid \psi_0 \rangle = 0$.
The second order correction to the energy is given by
\begin{equation}
	E^{(2)} = \sum_{ex} \frac{|\langle \psi_{ex} \mid H_1 \mid \psi_0 \rangle |^2}
	{\mathcal{E}_{ex}^{(0)} - \mathcal{E}_{gs}^{(0)}}.
\end{equation}
We can obtain the excited states that connect to the ground states $\mid \psi_0 \rangle$
by operating each term in $H_1$ on $\mid \psi_0 \rangle$; the resulting state
will be an excited state of $H_0$, whose unperturbed energy is computed by acting on this 
resulting $\mid \psi_{ex} \rangle$ by $H_0$. For example, consider the first 
exchange interaction term in $H_1$ which is $J_2 (\vec{S}_{1} \cdot \vec{S}_{3})$. 
When this operates on $\mid \psi_0 \rangle$, we get the state
\begin{eqnarray}
	J_2 (\vec{S}_1 \cdot \vec{S}_3) \mid \psi_0 \rangle &=& -\frac{J_2}{2 \sqrt{2}} \mid \psi_{ex} \rangle \nonumber \\
	& = & -\frac{1}{2\sqrt{2}} \mid \uparrow_1 \ \uparrow_2 \ \downarrow_3 \ 
	\overline{4 \ 5} \ \widefrown{6 \ \uparrow_7 \ 8} \rangle.
\end{eqnarray}
The unperturbed energy of the $\mid \psi_{ex} \rangle$, $\mathcal{E}_{ex}^{(0)}$ is 
given by $-\frac{J_1}{2} -\frac{3}{4} J_2 - B$. Similarly, we can calculate
the matrix elements of other exchange operators occurring in $H_1$, on the basis
of the eigenstates of $H_0$. This gives the second order corrected energy of 
the ground state of $H_0$ as
\begin{eqnarray}
	\mathcal{E}_{gs}^{(2)} &=& -\frac{3 J_1}{2} - \frac{3 J_2}{4} -\frac{9 J_2^2}{16 J_1} 
	\nonumber \\
	& & \qquad - \frac{J_2^2}{4(J_1+J_2)} - B.
	\label{eq:ene-gs}
\end{eqnarray}
Exact diagonalization of the skewed ladder Hamiltonian of 24 sites, for  
$J_1=5.0$ and $J_2=1.0$ gives a per site energy of  $-1.067$ as compared with
the perturbation theory prediction of $-1.05$ corresponding to an error of
$\sim 1.6 \%$.  Similarly the exact and second order corrected ground state
energy per site from perturbation theory, for $J_1=2.5$  and $J_2=1.0$ is
$-0.63$ and $-0.60$ respectively. Thus the error in perturbation theory 
is less than $5.0 \%$ in this case as well.  

Our numerical results show that the next plateau occurs at $m=1/2$, 
corresponding to the breaking of the (6, 8) singlet bond. The eigenstate
of the unperturbed Hamiltonian in this case is $\mid \overline{1 \quad 2} \ \uparrow_3
\ \overline{4 \quad 5} \ \uparrow_6 \uparrow_7 \uparrow_8 \rangle$.
The energy of the state correct to second order in perturbation is given by
\begin{eqnarray}
	\mathcal{E}_{m=1/2}^{(2)} &=& -\frac{3J_1}{2} + \frac{J_2}{4} - \frac{13J_2^2}{16J_1} -2B.
	\label{eq:ene-m1by2}
\end{eqnarray}
The error in energy per site of perturbation theory compared with exact results for 
24 sites in the absence of an applied magnetic field is $< 0.5 \%$ for $J_1 = 5.0$ and
$< 0.6 \%$ for $J_1 = 2.5$. From our numerical studies the ground state with $m=3/4$ 
is obtained by creating a triplet superposition of the two states formed 
by nearest-neighbor singlets and triplets from
spins at sites 1, 2, 4 and 5 while all other spins have $m_S = +1/2$ in the unit cell.
Thus we consider the $\mid \psi_{m=3/4}^{(0)} \rangle$ as given by
\begin{eqnarray}
	\mid \psi_{m=3/4}^{(0)} \rangle & = & \frac{1}{\sqrt{2}} \left[
		\mid \overline{1 \quad 2} \ \uparrow_3 \overrightarrow{4 \quad 5} \ 
        \uparrow_6 \uparrow_7 \uparrow_8 \rangle \right. \nonumber \\
	& & \qquad  
	\left. + \mid \overrightarrow{1 \quad 2} \ \uparrow_3 \overline{4 \quad 5} \ 
	\uparrow_6 \uparrow_7 \uparrow_8 \rangle \right],
\end{eqnarray}
where $\mid \overrightarrow{i \quad j} \rangle$ corresponds to $\mid \uparrow_i \uparrow_j \rangle$,
which is an $M_S = 1$ triplet. Using this as the unperturbed state we can obtain
the ground state energy of the unit cell in the $m=3/4$ state as
\begin{eqnarray}
        \mathcal{E}_{m=3/4}^{(2)} &=& -\frac{J_1}{2} + \frac{3 J_2}{4} - \frac{3J_2^2}{8J_1} - 3B.
	\label{eq:ene-m3by4}
\end{eqnarray}
This again has an error of $\sim 1 \%$ for $J_1 = 5.0$ and $< 5 \%$ for $J_1 = 2.5$
when compared with exact diagonalization results for the 24-site skewed ladder under a
periodic boundary condition in zero external field. In the fully polarized case, 
which corresponds to 
\begin{eqnarray}
        \mid \psi_{m=1}^{(0)} \rangle & = & \mid \uparrow_1 \uparrow_2 \uparrow_3
	\uparrow_4 \uparrow_5 \uparrow_6 \uparrow_7 \uparrow_8 \rangle,
\end{eqnarray}
the exact energy is trivially given by 
\begin{eqnarray}
	\mathcal{E}_{m=1} &=& \frac{J_1}{2} + 2 J_2 - 4B.
	\label{eq:ene-m1}
\end{eqnarray}

From the perturbation calculation, we obtain the critical fields for the onset 
of the plateaus at $m=1/2$, 3/4 and 1 which also correspond to the end of 
$m=1/4$, 1/2 and 3/4 plateaus. The on-set magnetic field for the first plateau, 
$B_{c_1}$, is obtained when the ground state with $m=1/4$ becomes degenerate 
with the lowest-energy state with $m=1/2$. Thus, by equating the r.h.s. of 
Eq.~\ref{eq:ene-gs} and Eq.~\ref{eq:ene-m1by2}, we get 
\begin{eqnarray}
	B_{c_1} &=& J_2 - \frac{J_2^2}{4 J_1} + \frac{J_2^2}{4(J_1 + J_2)}.
\end{eqnarray} 
Similarly, by equating the r.h.s. of  Eq.~\ref{eq:ene-m1by2} and Eq.~\ref{eq:ene-m3by4} we get $B_{c_2}$,
and by equating the r.h.s. of  Eq.~\ref{eq:ene-m3by4} and Eq.~\ref{eq:ene-m1} we get $B_{c_3}$ as
\begin{eqnarray}
	B_{c_2} &=& J_1 + \frac{1}{2} J_2  + \frac{7 J_2^2}{16 J_1}, \\
        B_{c_3} &=& J_1 + \frac{5}{4} J_2  + \frac{3 J_2^2}{8 J_1}.
\end{eqnarray}
From these critical fields we calculate the plateau widths as $w_1 = B_{c_1}$,
$w_2 = B_{c_2} - B_{c_1}$, and $w_2 = B_{c_3} - B_{c_2}$. These plateau widths 
are in good agreement with the numerical results, as seen in 
Fig.~\ref{fig:width} (b), for $J_1 > 1.5$.

\section{\label{sec:sum}Summary}
In this paper we have studied the magnetic properties of an antiferromagnetically 
interacting spin-1/2 system arranged on a 5/7 skewed ladder lattice 
(Fig.~\ref{fig:ladders}(b)). This system shows a high-spin gs with $m=1/4$ even 
in the absence of a magnetic field $B$, for $J_1 >2.35$;~\cite{geet} $J_1$ is 
the rung interaction, while the interaction between the nearest neighbors on 
the leg $J_2$ is set to $1$. This ladder system is also interesting and unique 
as it exhibits many plateaus. 
We have obtained $m-B$ curve, and we find that  there are three magnetization 
plateaus as a function of $B$. These three plateaus are at $m=1/4$, 1/2 and 
3/4,  consistent with the OYA condition. In the gs at $B=0$, each unit cell has 
three singlet dimers and  two ferromagnetically arranged free spins
(Fig.~\ref{fig:schematic}(a)). We find that each  plateau formation corresponds 
to successive breaking of a singlet dimer (Fig.~\ref{fig:schematic}). We have 
analytically obtained the widths of these plateaus as a function of $J_1$ from 
a simple perturbation theory. 

The plateaus at $m= 1/4$ and 3/4 appear even for small $J_1$ ($>0.4$), but the
$m=1/2$ plateau appears only for $J_1>0.8$. The width $w_i$ of the $i^{th}$ 
plateau at $m_i$ represents the magnitude of the energy gap in the system for that
particular magnetization. We notice that $w_i$ always increases with $J_1$; 
however, for $m=1/4$ and 1/2  it depends weakly on  $J_1$, and  at $m= 3/4$ the 
plateau width exhibits an almost linear variation with $J_1$. This is consistent with
the perturbation theory results. As usual, this system also shows cusps in $m-B$ 
curve at the beginning and end of the plateaus, and it follows the square-root 
dependence, $m \propto (B-B_c )^{1/2}$. The stability of the magnetization
plateaus in the presence of thermal fluctuation is an important factor for its 
observation. We notice that the plateaus at $m=1/4$ and 1/2 are robust against 
small thermal fluctuation,  while the plateau at 3/4 survives only up to 
$T/J_2 = 0.05$. The skewed ladder system can  be  mapped to a molecular system 
corresponding to fused five- and seven-membered carbon rings.~\cite{thomas2012}
Such a system corresponds to a fused azulene lattice,~\cite{thomas2012} and
it may be engineered at the grain boundary of a graphene sheet.~\cite{graphene_defect_nature_2011, arindam_ghosh_2015, balasubramanian2019}

\begin{acknowledgments}
	MK thanks Department  of Science and Technology (DST), India 
	for Ramanujan fellowship. SR thanks INSA Senior Scientist and 
	DST-SERB grant. 
\end{acknowledgments}

\bibliography{refplateau}
\end{document}